 \documentclass[preprint2]{aastex}

\slugcomment{submitted to the Astrophysical Journal}

\shorttitle{Helium shell flashes}
\shortauthors{Kato \and Hachisu}

\begin{document}


\title{Mass accumulation efficiency in helium shell flashes 
for various white dwarf masses}


\author{Mariko Kato}
\affil{Department of Astronomy, Keio University, Hiyoshi, Yokohama
223-8521, Japan}
\email{mariko@educ.cc.keio.ac.jp}

\and

\author{Izumi Hachisu}
\affil{Department of Earth Science and Astronomy, College of Arts and 
Sciences, University of Tokyo, Komaba, Meguro-ku, Tokyo 153-8902, Japan}
\email{hachisu@chianti.c.u-tokyo.ac.jp}



\begin{abstract}
We have calculated the mass accumulation efficiency 
during helium shell flashes on  white dwarfs (WDs) of mass 
0.7, 0.8, 0.9, 1.0, 1.1, 1.2, 1.3, and $1.35 ~M_{\odot}$
for the helium accretion rates 
$\log\dot M_{\rm He} ~(M_{\odot}$~yr$^{-1}) = -7.4$ to $-5.8$. 
This efficiency is a crucial factor for binary evolutions 
of Type Ia supernovae. 
For less massive WDs ($< 0.8 ~M_{\odot}$) no wind mass 
loss occurs and all the accreted mass accumulates on the WD
if the Roche lobe size is large enough. 
The efficiency takes the minimum values in between 
1.1 and 1.2 $M_{\odot}$ WD for a given mass accretion rate
and increases in both less and more massive WDs. 
The mass accumulation efficiency is larger 
than 0.5 for $\log\dot M_{\rm He} \geq -6.72$
in all the WD masses. 
\end{abstract}



\section{Introduction}

     Recent findings on the acceleration of the universe are
based on the brightness of Type Ia supernovae 
\citep[e.g.,][]{per99, rie98}.  
However, the diversity of the brightness
has not been fully understood yet, which may be
related to their binary evolution paths \citep[e.g.,][]{ume99}.
In order to elucidate the physics of Type Ia supernovae, 
every evolutionary path to Type Ia supernovae should be examined, 
even though its production rate is very rare.  
In these paths, the response of white dwarfs (WDs) to hydrogen/helium
matter accretion is a key process and, especially, the efficiency of
mass accumulation is crucial.

     Recent binary evolution scenarios of Type Ia supernovae
include a WD accreting matter lost by a companion
\citep*{hac96, hac99a, hac99b, han04, lan00, li97},
although the merger model of double degenerates has not been fully
rejected yet \citep[e.g.,][]{liv03}.
In these scenarios, the WD  grows in mass and explodes as a
Type Ia supernova when it reaches 
the Chandrasekhar mass limit.   When the accretion rate of hydrogen 
is much smaller than $\sim 1 \times 10^{-7} M_\sun$~yr$^{-1}$,
hydrogen shell flash triggers a nova outburst.  Almost all of 
the accreted matter is blown off or even the WD is eroded
by convective dredge-up \citep[e.g.,][]{pri86}.
When the accretion rate is as large as 
$\sim 1 \times 10^{-7} M_{\odot}$ yr$^{-1}$ or more, the shell flash 
is relatively weak or hydrogen burning is stable so that
most part of the accreted matter is processed to helium 
and then accumulates on the WD.  

Another possible way to Type Ia supernovae is an helium accretor;
the WD accretes helium from companion helium star and grows in mass to 
the Chandrasekhar mass limit \citep[e.g.,][]{yoo03}.
In such a case, a helium shell flash occurs to develop 
a nova outburst when the mass accretion rate is smaller than a critical
value, which varies from $1 \times 10^{-6}$ to 1 $\times 10^{-5} 
M_\sun$~yr$^{-1}$, depending on the WD mass.  
If the helium shell flashes occur, the wind mass loss carries away 
a part of the envelope matter, which reduces the growth rate of the WD.
\citet{yoo03} have calculated binary evolution that contains a helium
accreting WD. In their calculation, the effect of wind mass loss 
is partly taken into account based on an empirical formula of 
Wolf-Rayet star mass loss, whereas
\citet{han04} have used the mass accumulation efficiency of 
helium accretion  calculated by \citet{kat99} for a 1.3 $M_{\odot}$ WD. 
In this way, the mass accumulation efficiency is a crucial factor for 
evolution of the binary.

\citet{iva04} have calculated evolution of WD binaries in a
phase of thermal timescale mass-transfer.  The crucial factor 
of the binary evolutions is the reaction of WDs against 
high mass transfer rate, that is, the mass accretion efficiency
for hydrogen/helium shell flashes.
\citet{taa04} addressed the mass accumulation efficiency
of helium novae is one of the three fundamental factors
that should be clarified in order to advance the evolutionary
scenarios to the next step.  The mass accumulation efficiency
for helium shell flashes has been calculated but 
only for a $1.3 ~M_{\odot}$ WD \citep{kat99}. 

From the observational point of view, such a helium shell flash should
be observed as a helium nova outburst.  Recently, in late 2000,
V445 Puppis was discovered at its outburst stage. 
\citet{ash03} suggested, from its infra-red spectrum features, 
that V445 Pup is the first identified helium nova. 
\citet{kat03} calculated light curves of helium novae 
for various WD masses to fit them with the observational data.
They concluded that the light curve can be well fitted by
a helium nova on a WD having its mass more massive than
$1.33 ~M_\sun$.  This massive WD indicates that the WD is now
growing in mass even for such a violent helium shell flash.
\citet{kat03} suggested that V445 Pup is a progenitor of 
Type Ia supernova.
Therefore, we are forced to realize the importance of 
mass loss/mass accumulation during helium shell flashes 
on the WD.

\citet*{kat89} calculated the evolution of helium shell flashes
on an $1.3 ~M_\sun$ WD.  \citet{kat99} have recalculated the 
helium shell flashes with OPAL opacity 
and obtained the mass accumulation efficiency, i.e., 
the ratio of the envelope mass accumulated after one cycle 
of helium shell flash to the ignition mass.  
These two works are done only for a $1.3 ~M_\sun$ WD 
and no results have been presented for other WD masses.
If the accumulation efficiency is very 
small for less massive WDs, the evolution scenarios should be 
drastically changed.  In this Letter, therefore, we have calculated  
mass accumulation efficiency of helium shell flashes 
for various WD masses.

 
\section{Input Physics}
The decay phase of helium nova outbursts is followed 
based on an optically thick wind theory \citep{kat94}.  
The structure of the
WD envelope is calculated by solving the equations of motion, 
continuity, energy transport, and energy conservation. 
The details of computation have been already presented 
in \citet{kat99} for helium shell flashes on a $1.3~M_\sun$ WD.

Here, we examine other various WD masses, i.e., 
0.7, 0.8, 0.9, 1.0, 1.1, 1.2, and $1.35~M_\sun$.  
The radius of the WDs are assumed 
as given in Table 1 because of the following reasons:
In time-dependent calculation \citep{kat89},
the radius of the nuclear 
burning shell moves back and forth during one cycle of a 
helium shell flash.  In the $1.3 ~M_\sun$ WD case, 
we assumed the radius as a time-averaged mean value of 
the radius \citep[see, e.g.,][]{kat89, kat99}. 
For WD masses other than $1.3~M_{\odot}$, however,
no time-dependent calculation of helium shell flashes is presented.   
Therefore, we estimate the radius from an geometrical mean 
of the two radii, i.e., the naked C+O core radii of two
steadily-accreting WD with the mass accretion rate of 
$\dot M_{\rm He} = 1.0 \times 10^{-7}$ and $1.0 \times 10^{-6}
 ~M_{\odot}$~yr$^{-1}$(private communication with H. Saio 2003). 

After the onset of a shell flash, 
helium is processed to carbon and oxygen. 
Helium burning produces nuclear energy, 
a large part of which is consumed 
for the envelope matter to be push 
upward against the gravity. 
The nuclear energy release is almost comparable 
to the gravitational potential, and a large part of 
the envelope is processed 
in very early stage of explosion, i.e., before the 
expansion of the envelope. The processed 
matter is then mixed into the whole envelope by convection.
Therefore, we assume the composition of the envelope 
to be uniform with the values of $Y$ and $C+O$ as listed in 
Table 1. Here, we assume $X=0.0$, and $Z =0.02$, 
where $Z$ includes $C$ and $O$ in the solar abundance ratio.
The value of $C+O = 0.5$ for the $1.3 ~M_\sun$ WD 
is taken from \citet{kat99}. 
Changing the ratio of carbon to oxygen with the total mass ratio
constant ($C+O = $const.) hardly changes the result.
OPAL opacity is used.

\section{Mass Accumulation Efficiency}

Figure 1 shows the photospheric temperature $T_{\rm ph}$, 
the photospheric wind velocity $V_{\rm ph}$, the photospheric 
radius $R_{\rm ph}$, the wind mass loss rate  $\dot M_{\rm wind}$ ({\it dotted}),
and the total mass decreasing rate of the envelope  $\dot M_{\rm tot}= 
\dot M_{\rm wind}+ \dot M_{\rm nuc}$, ({\it solid}), here the 
mass decreasing rate owing to nuclear burning is $\dot M_{\rm nuc}=L_
{\rm nuc}/(\epsilon_{\rm He}Y)$, where $L_{\rm nuc}$ is the nuclear 
luminosity.
The solution of $1.3~M_\sun$ is already presented in \citet{kat99}.
At the maximum expansion of the envelope after the onset 
of a shell flash, the star reaches 
somewhere on the curve, depending on the envelope mass
at ignition, and moves leftward in time. The wind mass 
loss stops at the point marked by the small open circles. 
After that, the star continues to move leftward 
owing to nuclear burning.  
Helium nuclear burning stops at the end of each curve. 
In the case of $0.8~M_\sun$ WD, we have obtained 
only a very short sequence, because of a numerical difficulty 
\citep[see][for more details]{kat94}. 
No winds occur in the $0.7~M_\sun$ WD. 

Figure 2 shows the mass accumulation efficiency 
$\eta_{\rm He}$ defined by the ratio 
of the processed matter remaining after one cycle 
of helium shell flash to the ignition mass. The amount of matter processed 
or lost by the wind are calculated from the wind mass loss rate 
and nuclear burning rate in Figure 1 \citep[see][for details]{kat89}. 
Here, we use the relation between the ignition mass and the helium
accretion rate (private communication with H. Saio 2004) obtained based 
on a steady state in which non-homologous term of gravitational energy 
generation is neglected \citep{nom82}.
For the case of $0.7~M_\sun$, no wind occurs and all of the envelope
matter is processed to accumulate on the WD, if the binary separation 
is large enough for the expanded envelope to reside in the Roche lobe.
In low accretion rates ($\log \dot M_{\rm He}(M_{\odot}$yr$^{-1}) < -7.6$) a helium 
detonation may cause a supernova explosion.

The value of $\eta_{\rm He}$ in Figure 2 is approximated  by the following 
algebraic form.

\begin{equation}
{\eta_{\rm He}}= \cases{  -0.115 ({\log \dot M_{\rm He}}+ 5.7)^2 + 1.01 ,
          \cr ~~~~~~(-7.4 < \log \dot M_{\rm He} < -6.05 )
        \cr 1,   ~~~~~~ ( -6.05 \leq \log \dot M_{\rm He}) \cr}
\end{equation}

\noindent
for 1.35 $M_{\odot}$ WD,

\begin{equation}
{\eta_{\rm He}}  = \cases{ -0.175 ({\log \dot M_{\rm He}}+ 5.35)^2 + 1.03 ,
          \cr ~~~~~~( -7.35 < \log \dot M_{\rm He} < -5.83)
         \cr 1, ~~~~~ ( -5.83 \leq \log \dot M_{\rm He}) \cr}
\end{equation}

\noindent
for 1.3 $M_{\odot}$ WD,

\begin{equation}
{\eta_{\rm He}}  = \cases{ 0.54 {\log \dot M_{\rm He}} + 4.16
        \cr ~~~~~~~~ ( -7.06 < \log \dot M_{\rm He} < -5.95 )\cr
        -0.54 ({\log \dot M_{\rm He}}+ 5.6)^2 + 1.01 ,
       \cr ~~~~~~( -5.95 \leq \log \dot M_{\rm He} < -5.76 ) \cr
         1,  ~~~~~~~~~( -5.76   \leq \log \dot M_{\rm He}) \cr}
\end{equation}

\noindent
for 1.2 $M_{\odot}$ WD. This equation is well fitted also to
1.1 $M_{\odot}$ WD.

\begin{equation}
{\eta_{\rm He}}  = \cases{-0.35 ({\log \dot M_{\rm He}}+ 5.6)^2 + 1.01 ,
        \cr  ~~~~~~~(  -6.92 < \log \dot M_{\rm He} < -5.93) \cr
          1,  ~~~~~~~~~~~~( -5.93 \leq \log \dot M_{\rm He})\cr}
\end{equation}

\noindent
for 1.0 $M_{\odot}$ WD,

\begin{equation}
{\eta_{\rm He}}  = \cases{-0.35 ({\log \dot M_{\rm He}}+ 5.6)^2 + 1.07 ,
         \cr  ~~~~~~( -6.88 < \log \dot M_{\rm He} < -6.05) \cr
          1,   ~~~~~~ ( -6.05 \leq \log \dot M_{\rm He})\cr}
\end{equation}

\noindent
for 0.9 $M_{\odot}$ WD,

\begin{equation}
{\eta_{\rm He}}  = \cases{-0.35 ({\log \dot M_{\rm He}}+ 6.1)^2 + 1.02 ,
          \cr  ~~~~~~~~(-6.5 < \log \dot M_{\rm He} < -6.34) \cr
          1,   ~~~~~~~( -6.34 \leq \log \dot M_{\rm He})\cr}
\end{equation}

\noindent
for 0.8 $M_{\odot}$ WD.

When the WD mass is increased, winds begin to blow 
at $\sim 0.8~M_\sun$.  The winds become stronger
as the WD mass is further increased. 
The accumulation efficiency $\eta_{\rm He}$ then decreases 
because of much stronger wind mass loss.
The 1.1 and $1.2~M_\sun$ WDs take the smallest values 
of $\eta_{\rm He}$ for all accretion rates.
For more massive WDs, $\eta_{\rm He}$ increases with the WD mass
because the nuclear burning rates are much higher
than the wind mass loss rates owing to the strong gravity.
As a result, most of the envelope mass accumulates on these
more massive WDs.

\section{Discussion and conclusions}

The accumulation efficiency has different values if we assume
a different chemical composition or a WD radius.
As the $C+O$ content becomes larger, the mass accumulation ratio 
increases because the wind becomes weak but the nuclear burning rate
increases.  For example, if we increase $C+O$ from 0.5 to 0.6 
in the $1.3 ~M_\sun$ case, $\eta_{\rm He}$ increases by 0.08, 
and from 0.3 to 0.4 in the $1.2 ~M_\sun$ WD,
$\eta_{\rm He}$ increases by $0.04-0.05$. 
If we assume a smaller WD radius, the wind becomes stronger, 
and we get a smaller accumulation efficiency $\eta_{\rm He}$. 
For example, if we take the Chandrasekhar radius 
for the $1.2~M_\sun$ WD as the smallest limit, 
$\eta_{\rm He}$ decreases by 
$0.07$  at $\log\dot M_{\rm He} ~(M_\sun$~yr$^{-1}) = -6.8$ 
and by 0.12 at $\log\dot M_{\rm He} ~(M_\sun$~yr$^{-1}) = -6.0$. 

For the $0.6~M_\sun$ WD, no wind occurs.  All the accreted mass 
is processed to carbon and oxygen and accumulated on the WD. 
We have not presented solutions of the $0.6~M_\sun$ WD in Figure 1,
because the WD radius changes largely depending on 
the mass accretion rate between 
$\log\dot M_{\rm He} ~(M_\sun$~yr$^{-1}) = -7$ and $-6$.  In addition, 
a density inversion layer appears around  
$\log T ~({\rm K}) \sim 5.2$, which 
causes a numerical difficulty of the computation.

  We have assumed no dredge-up of WD material into the envelope.
If WD material under the accreted helium layer is mixed into the envelope,
the helium content $Y$ is reduced, which increases  the mass 
decreasing rate owing to nuclear burning becomes larger, because $\dot M_{\rm nuc}
\propto Y^{-1}$. On the other hand, the wind mass loss becomes weak due to the decrease
in opacity, i.e., $\dot M_{\rm wind}$ becomes smaller. Therefore, mixing of WD 
matter has effects to increase the mass accumulation efficiency, because
$\eta \sim  \dot M_{\rm nuc}/(\dot M_{\rm nuc} + \dot M_{\rm wind})$.

For example, in the case of $1.3 M_{\odot}$, if the same amount of WD mass is
mixed to the envelope mass, i.e., $Y$ is $\sim 0.5$ at ignition.  It reduced
much more in the course of rising phase, because a time dependent calculation
(Kato, Saio  and Hachisu 1989) shows that $Y$ decreases from 0.98 to 0.5 in the
rising phase. As the gravitational potential is comparable to nuclear energy
generation as shown in table 1, a large part of helium is consumed to lift
the whole envelope up against the gravity at the beginning phase of outburst.
Therefore, $Y$ is reduced to $\sim 0.1$ or less. The change of $Y$ causes the change
in nuclear burning rate and the opacity. As a results,
the wind mass loss does not occur for the envelope less than $1.5\times 
10^{-4} M_{\odot}$,
which correspond to the mass accretion rate $\log \dot M (M_{\odot}$yr$^{-1})= 
-6.51 $ i.e., $\eta_{\rm eff}$ is almost 100 percent at $\log \dot M> -6.51$.
Even the wind occurs, the mass loss rate is as low as 1/10 of nuclear burning
rate and almost all of the envelope mass will accumulate on the WD.

  It is not known whether WD material is dredged up or not, or how much
amount is dredged up. In the case of hydrogen shell flashes, enrichment of
WD material is observed in ejecta of classical novae, which is an evidence
of dredged-up of WD material. However, no heavy element enrichment is
observed in recurrent nova ejecta. Therefore, we may conclude that
the dredge up mechanisms do not effectively work if the recurrence period
is short (or the mass accretion rate is high).

In the case of He novae, only one object, V445 Pup, is observed so far.
There are many bright C lines in outburst phase, but we don't know this
carbon is originated from WD material or processed helium. The WD mass
is estimated to be $\geq 1.33 M_{\odot}$, and the recurrence period to be
several tens of years (Kato and Hachisu 2003). Such a very massive WD
itself may suggest that the WD is now growing, but not that dredge up
mechanisms are working.

\citet{cas98} have criticized Hachisu et al.'s (1996) model
in which accreting WDs grow in mass and reach the Chandrasekhar 
mass limit.  The main point of their criticism is that \citet{hac96} 
have neglected the interaction between the donor and the extended 
envelope during helium shell flashes.  \citet{cas98} speculated 
that most of the envelope mass is lost from the binary by the effect
of the companion and, as a result, the WD cannot grow from 
$\sim 1 M_{\odot}$ to the Chandrasekhar mass limit.  
If so, the Type Ia scenario proposed by \citet{hac96} is seriously
damaged.  However, as already discussed in \citet{kat91a, kat91b, kat94},
the viscous heating by the companion motion in the envelope is 
not effective when the wind blows.  It is because the wind 
velocity is much faster than the companion motion and the wind 
quickly escapes from the binary system with almost no interaction
between them.  For WDs more massive than $0.8~M_\sun$, 
the wind is fast enough and therefore we can neglect effects of 
interaction between the wind and the companion's orbital motion.
Once the optically thick wind blows and its velocity is faster than
the orbital motion, we are able to estimate the accumulation efficiency 
only from the wind solutions.

\citet{cas98} adopted Los Alamos opacity that shows much smaller
values (about one-third) than those of OPAL opacity around 
the temperature of $\log T \sim 5.2$. 
The envelope solutions for Los Alamos opacity are largely different from
for OPAL opacity when the photospheric temperature is lower than
this value (i.e., $\log T_{\rm ph}$ (K)$ < 5.2)$.
When OPAL opacity is adopted, 
we obtain a more massive envelope mass and a higher 
temperature at the hydrogen burning region 
for the same photospheric radius.
As a result, we have a higher nuclear burning rate 
for a higher temperature. 
A larger part of the envelope mass is quickly processed into $C+O$ and 
accumulates on the WD.  
Therefore, we conclude that a higher mass accumulation efficiency
is resulted even when the strong wind blows.

Our main results are summarized as follows: 

The mass accumulation efficiency of helium shell flashes,
$\eta_{\rm He}$, is a crucial factor for binary evolutions 
of Type Ia supernovae.  For WDs more massive than 
$0.8 ~M_\sun$, helium shell flashes trigger an optically thick
wind mass loss.  We have calculated $\eta_{\rm He}$ 
for $0.8 - 1.35 ~M_\sun$ WDs; more than 50\% ($\eta_{\rm He} > 0.5$)
is accumulated on the WD after one cycle of helium shell flash
for $\log\dot M_{\rm He} ~(M_\sun$~yr$^{-1}) \geq -6.72$ 
on the WD masses of $0.8 - 1.35 ~M_\sun$.
For WDs less massive than $0.7~M_\sun$,
no wind mass loss occurs and all the accreted matter 
remains on the WD if we can neglect the effect of the Roche lobe overflow.

\begin{deluxetable}{llcll}
\tabletypesize{\scriptsize}
\tablecaption{Model parameters. \label{tbl-1}}
\tablewidth{0pt}
\tablehead{
\colhead{$M_{\rm WD}$} & \colhead{$\log R_{\rm WD}$} 
& \colhead{$GM_{\rm WD}/R_{\rm WD}^2 q$} & \colhead{${C+O}$} & \colhead{$Y$}\cr
\colhead{($M_{\odot}$)} & \colhead{($R_{\odot}$)} &  &   &  
}
\startdata
0.7  &-1.74  & 0.12  &0.06 &0.92\cr
0.8  &-1.84  & 0.17  &0.1  &0.88\cr
0.9  &-1.93  & 0.24  &0.13 &0.85\cr
1.0  &-2.01  & 0.32  &0.2  &0.78\cr
1.1  &-2.09  & 0.43  &0.23 &0.75\cr
1.2  &-2.19  & 0.60  &0.3  &0.68\cr
1.3  &-2.33  & 0.88  &0.5  &0.48\cr
1.35 &-2.44  & 1.18  &0.7  &0.28
\enddata
%
\end{deluxetable}

\begin{figure}
\plotone{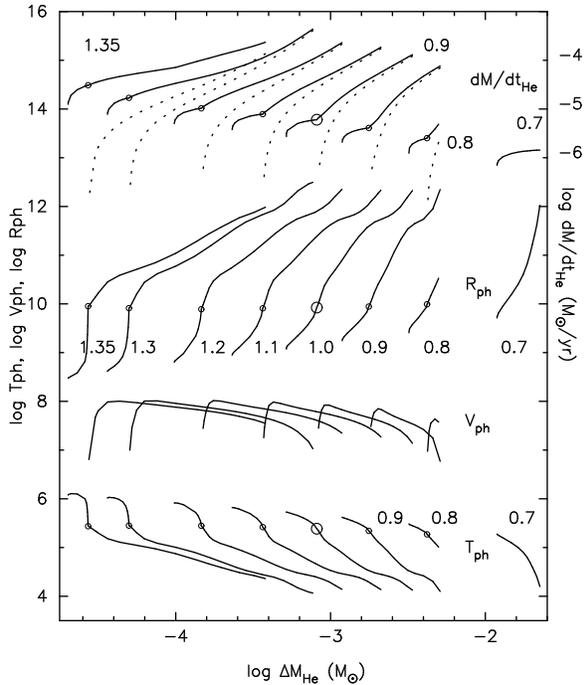}
\caption{ {\it Top}: Total envelope mass decreasing rate, i.e., 
nuclear burning + wind mass loss ({\it solid curve}),
and wind mass loss rate ({\it dashed curve}) 
in units of $M_\sun$~yr$^{-1}$, 
{\it Second}: Photospheric radius $R_{\rm ph}$~(cm), 
{\it Third}: wind velocity $V_{\rm ph}$~(cm~s$^{-1})$, 
{\it Bottom}: temperature $T_{\rm ph}$~(K), against
the helium envelope mass, $\Delta M_{\rm He}$, in units of $M_\sun$. 
The wind mass loss does not occur in left side of the point marked by a 
small open circle (an enlarged circle for 1.0 $M_\sun$)
The WD mass is 1.35, 1.3, 1.2, 1.1, 1.0, 0.9, 0.8, and 0.7 $M_\sun$, 
from left to right.
\label{fig1}}
\end{figure}

\begin{figure}
\plotone{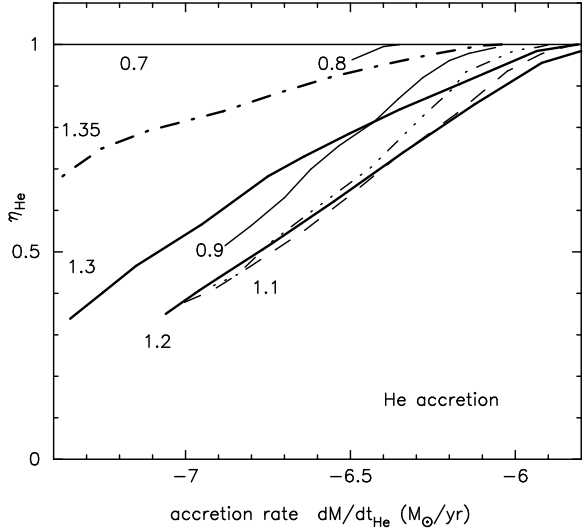}
\caption{ Mass accumulation efficiency, $\eta_{\rm He}$, is plotted 
against helium mass accretion rate for a $0.7 ~M_\sun$ WD
({\it horizontal solid line}), 0.8 and $0.9 ~M_\sun$ WDs
({\it thin solid line}), $1.0 ~M_\sun$ WD ({\it dash-three-dotted line}), 
$1.1 ~M_\sun$ WD ({\it dashed line}), 
1.2 and $1.3 ~M_\sun$ WDs ({\it thick solid line}), 
and $1.35 ~M_\sun$ WD ({\it dash-dotted line}).  
\label{fig2}}
\end{figure}
%
%
%
\end{document}